\documentclass{fundam}
\usepackage{url} 
\usepackage[ruled,lined]{algorithm2e}
\usepackage{graphicx}
\usepackage{amsfonts}
\usepackage{amssymb}

\begin{document}

\title{Iterative method of generating artificial context-free grammars}

\address{olgierd.unold@pwr.edu.pl}

\author{Olgierd Unold\\
Department of Computer Engineering \\
Wroc\l{}aw University of Science and Technology \\ Wybrzeże Wyspiańskiego 50-370 Wroc\l{}aw, Poland\\
olgierd.unold{@}pwr.edu.pl
\and Agnieszka Kaczmarek\\
Department of Computer Engineering \\
Wroc\l{}aw University of Science and Technology \\ Wybrzeże Wyspiańskiego 50-370 Wroc\l{}aw, Poland\\
agnieszka.kaczmarek{@}pwr.edu.pl
\and \L{}ukasz Culer\\
Department of Computer Engineering \\
Wroc\l{}aw University of Science and Technology \\ Wybrzeże Wyspiańskiego 50-370 Wroc\l{}aw, Poland\\
lukasz.culer{@}pwr.edu.pl  } \maketitle

\runninghead{O. Unold, A. Kaczmarek, \L{}. Culer}{Iterative method of generating artificial context-free grammars}

\begin{abstract}
Grammatical inference is a machine learning area, whose fundamentals are built around learning sets. At present, real-life data and examples from manually crafted grammars are used to test their learning performance. This paper aims to present a method of generating artificial context-free grammars with their optimal learning sets, which could be successfully applied as a benchmarking tool for empirical grammar inference methods.
\end{abstract}


\section{Introduction} \label{introduct}

The study area called grammatical inference contains a large variety of methods \cite{de2010grammatical}. Some of them require specific conditions to be fulfilled (such as a number of learning examples, assuming a certain type of grammar that generated those examples) but exactly learn the target grammar in a predictable time - they are called \emph{formal grammar induction methods}.  The other group, \emph{empirical grammar induction methods}, do not promise to provide a precise solution. However, they contain specific heuristics that, with the use of an example set, attempt to predict the grammar structure. They are usually developed for specific language families (e.g.. context-free languages) and their performance is tested in an experimental approach. There are two common sources of learning data for them - \emph{real-life data} (such as peptide sequences \cite{AmyloidDB}) or sets for \emph{manually crafted grammars} \cite{tomita1982dynamic}.

\emph{Real-life data} framework supplies algorithms with examples representing a rich variety of features - the high performance, achieved with those sets is promising for practical applications. However, they are consistently more demanding in terms of learning (some algorithms do not infer them properly), in addition, there is a general scarcity of knowledge about grammars which describe them - if they exist at all. Finally, examples for some of those learning sets are obtained in laboratories using physical equipment, which allows errors to occur in their sequences.

\emph{Sets from manually crafted grammars} have multiple advantages. Due to the fact that grammars are created by hands, we possess full knowledge of them. There is also a possibility to create learning sets, which could contain both positive and negative examples, free from errors in sequences. However, this is not a trivial task - some languages could contain infinite examples, so a complete overview is not possible for them. Selecting proper examples for them is a separate, complex issue. Reliable procedures exist for certain language families - e.g. for CFGs - presented in section \ref{regpositive} and \ref{regnegative}.  Finally, manually creating complex grammar with a wide spectrum of features is a difficult and time-consuming process, although there are some tools that simplify it \cite{rohde1999simple}.

One of empirical grammar induction method is \emph{GCS (the Grammar-based Classifier System)}, introduced in 2005 \cite{unold2005context}. It derives from \emph{LCS (the Learning Classifier System)} \cite{john1986holland} which combines two biological metaphors - evolution and learning. It develops a set of classifiers in a form of condition-action rules adapted to a set of positive and negative input examples. GCS, as an extension, implemented new features as, such as:
\begin{itemize}
  \item a new representation of classifier population,
  \item a new scheme of classifiers matching to the environmental state,
  \item new methods of exploring new classifiers.
\end{itemize}

One issue that concerned us during our research was testing the performance impact of new improvements. The currently existing learning sets have, naturally, fixed the complexity at a given level, which usually does not allow to easily notice subtle changes in algorithm performance, because of a few reasons:
\begin{itemize}
  \item the low complexity of some sets (for every learning set, the full performance was obtained anyway),
  \item  the high complexity of some sets (method was unable to learn grammar regardless of the new feature),
  \item a lack of specific features needed to test a given feature (performance was not changed at all).
\end{itemize}

There was a need to have something in the middle, where the “middle” was defined differently for each problem - having access to a wide range of learning sets with different levels of complexity and containing different types of features would allow one to easily match the proper set to highlight the impact of new improvements as much as possible. 

Unfortunately, no set category allows one to obtain those middle sets well. Creating a new grammar of given complexity for manually created ones requires precise and time-consuming planning and crafting. For real-life sets obtaining those sets of given parameters is nearly impossible - an approximated solution would require the selection of a specific type of examples, based on prior knowledge, observation and intuition, which is rarely possible.

To provide a solution for those difficulties, in this article we present a complete and unique approach for automatic generation of coherent grammars. It allows specifying required grammar complexity (using the defined measure) or a number of specific rules. Integrated additional modules allow also to create positive and negative learning sets for a given grammar. All of the mentioned constituents make the output tailored to specific research needs, without additional effort (grammar is generated by a computer, the user has only to define its needs) and time (few seconds of generation process) spent on creating a grammar and sets manually. We decided to rely on context-free languages due to the richness of theoretical background, which we utilized to reliably justify theoretically our approach, making it a solid base for grammar inference related research.

This paper will start with describing all components - from the Grammar Generator (section \ref{grammargen}), followed by the grammar complexity analysis (section \ref{gcisec}) and the Positive Set Generator (section \ref{regpositive}) to the Negative Set Generator (section \ref{regnegative}). Then all previously described modules will be connected in section \ref{reqflow}, which describes overall system requirements and flow. Finally, a test study will be performed (section \ref{teststudy}). In the end, a discussion (section \ref{conc}) with future development plans (section \ref{fwork}) will be included.

\section{Grammar generator}
\label{grammargen}

\subsection{Preliminaries} \label{prelimi}
This section will introduce some of basic grammar inference definitions. Let $G=(N, \Sigma, R, S)$ be a context-free grammar (\emph{CFG}). In presented quadruple $N$ is a finite set of non-terminal symbols, $\Sigma$ is a finite set of terminal symbols ($N \cap \Sigma = \emptyset $), $R$ is a finite set of production rules, $R\subset N\times(\Sigma\cup N)^*$, and $S\in N$ is the start symbol. Context-free grammar rules are presented in the for of $A \rightarrow \alpha$, where $A \in N$ and $\alpha \in (\Sigma \cup N )^\ast$. 

Any context-free grammar $G$ can be transformed into Chomsky Normal Form (CNF). CNF grammars has only two types of rules:
\begin{itemize}
  \item $A \rightarrow BC$,
  \item $A \rightarrow a$,
\end{itemize}
where $A,B,C \in N$ and $a \in \Sigma$. 

Let $L(G)$ denote the set of strings (the string language) generated by $G$ over the alphabet $\Sigma$, for which there exists a derivation $s \Rightarrow^\ast x$ using rules in $R$, where $x\in\Sigma^*$. The time complexity to recognize whether a certain sentence (string) belongs to a given context-free language $L(G)$ is $O(n^3)$, where $n$ is the length of a sentence.

\subsection{Idea, algorithm requirements and theoretical background}

The underlying idea of the grammar generator is to create an automatic procedure that allows one to create consistent, artificial context-free grammars. One of the required attributes was the ability to set generated grammar complexity smoothly. 

Firstly, the type of the created grammar and the form of its rules should be selected. We decided to utilize context-free grammars, because of the wide theoretical background, as was previously mentioned. One publication provides a basis for rule form selection - in \cite{sakakibara2005learning} three fundamental types of CFG rules are mentioned:

\begin{enumerate}
\renewcommand{\labelenumi}{(\alph{enumi})}
  \item Parenthesis rules - rules in the form of \emph{A $\rightarrow$ aBb $\lor$ A $\rightarrow$ab}. They are built using one non-terminal symbol on the left side, and two terminal symbols on the right side, with the possibility of one non-terminal symbol between them. In the cited publication the number of non-terminal symbols is potentially unlimited; however, we restricted it to one in order to simplify the model without losing description possibilities - the parenthesis rule with more than one non-terminal symbol on its right side can be easily converted to a set of multiple branch rules and one parenthesis rule with at most one non-terminal symbol on the right side.
\item Branch rules - rules in the form of \emph{A $\rightarrow$ CD}. They consist of one non-terminal symbol on the left side and two non-terminal symbols on the right side.
\item Iteration rules - rules in the form of \emph{A $\rightarrow$ cE $\lor$ A $\rightarrow$ Ec}. They consist of one non-terminal symbol on the left side and a non-terminal symbol with a terminal one in any order on the right side.
\end{enumerate}

Secondly, the internal requirements for grammar should be specified. Starting from the basics - we expect that grammar will not be a set of random rules. Instead, it would allow the creation of a non-empty set of words. Furthermore, we should avoid creating rules that do not have a real impact on generating words - there is no possibility for their presence in a derivation tree in any case. To achieve that, we need to describe additional concepts:

\emph{Rule / Symbol achievability} - we consider both a non-terminal or terminal symbol as achievable when there exists a path, created using grammar rules, that leads from the starting symbol to the examined one; this means that we can derive it from the starting symbol. Analogously, it can be defined for rules - it is considered achievable when there exists a path, created using grammar rules, that leads from the starting symbol to the examined rule. The start symbol is considered to be achievable by itself, due to the fact that it is used as the initial symbol \cite{hopc}.

\emph{Rule / Symbol productivity} - we consider a symbol as productive when at least one rule that derives from that symbol is productive. A rule is considered productive when all symbols that it generates are productive. Terminal symbols are productive, as they are the final stage of word derivation. Productivity, in terms of programming, is determined in an iterative manner starting from terminal symbols and labeling proper non-terminal symbols as productive in later steps \cite{hopc}.

\emph{Grammar consistency} - we consider a grammar to be consistent when all the rules and symbols that the examined grammar contains are achievable and productive. Creating a consistent grammar is the goal of our generation process.

Thirdly, external requirements (given by algorithm user) should be specified. The user inputs requirements in the form of parameters. There are two approaches to specifying those parameters. They can be passed explicitly, by setting each parameter at a specific value, or a target number of grammar rules is provided (the procedure of determining parameters based on it is described with details in section \ref{gcisec}).

At the very end, the determined parameters are verified using an equation system derived in section \ref{tanalysis}. As it will be proven, if an equation system completed with given parameters is consistent, the grammar is feasible to be generated. Otherwise, the user is notified (when parameters are set directly) or parameters are drawn afresh.

\subsection{The algorithm}

\begin{figure}[ht!]
\begin{center}
\includegraphics[width=0.8\linewidth]{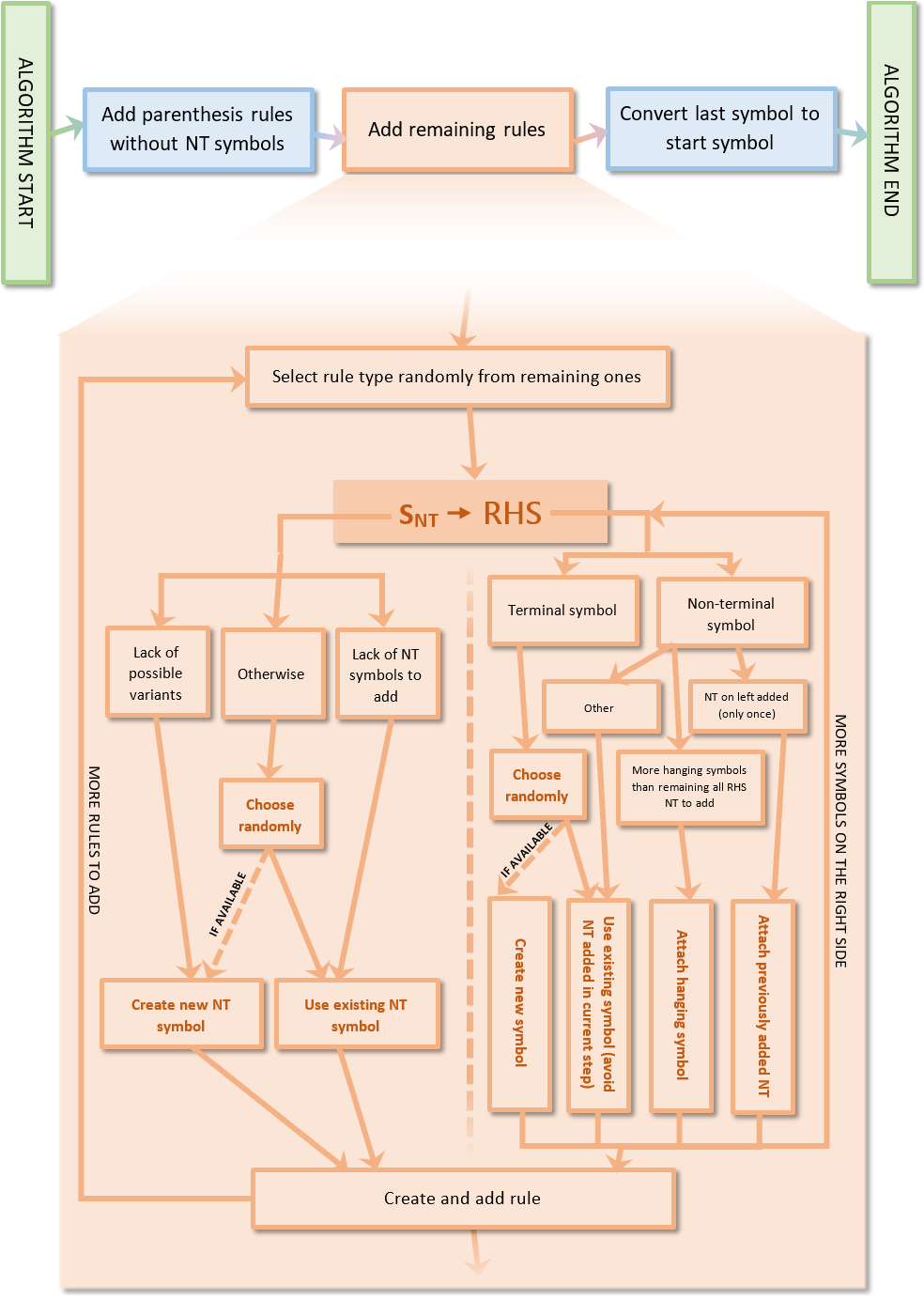}
\caption{Grammar generation algorithm illustrated with a diagram}
\label{algorithm}
\end{center}
\end{figure}

The algorithm is based on an iterative approach. Due to high complexity, algorithm pseudocode with detailed instructions was replaced by a more general diagram (Fig. \ref{algorithm}) with description.

Initially, rule and symbol sets are empty. Rules are added until the requirements, defined in the previous section, are fulfilled. The general idea of the algorithm is adding all demanded rules iteratively with the respect to the claimed principle, that all added rules and symbols had to be productive just after adding them to the grammar. Due to that, firstly all parenthesis rules without non-terminal symbols are added, since they are productive by themselves, without the need of connection to any other rules. Symbols used to create them are chosen randomly (can be both shared or exclusive to the given rule) having regard to limitations expressed as a system of inequalities (\ref{EQ:9}), which would be explained and proved in the following sections. All, except for the lastly added non-terminal symbol created during this step will be referenced as "hanging" symbols.

The most complex part, highly dependent on the current state of grammar, is adding the next rules. As previously, the whole process is performed according to defined principles:
\begin{itemize}
  \item new rule type is selected randomly from available ones. Symbols, for both left and right side, could be picked from currently existing ones or created, if this would not violate following principles and the system of inequalities (\ref{EQ:9}), 
  \item rules are being connected only to productive symbols (only productive symbols have to be on the right side of the rule - new non-terminal symbol cannot be used),
  \item rules can be added using the existing non-terminal symbols on their left side or by creating new ones,
  \item if a new non-terminal symbol is created, one of the non-terminal symbols from the right side has to be a recently added non-terminal symbol - that rule ensures, that all currently connected symbols would be achievable from created symbol (except those, that are considered as hanging)
  \item the new non-terminal symbol cannot be attached to the right side of a rule, for which it was created to avoid cycles creation,
  \item all rules have to be unique - there should not exist any pair of rules, that have both left and right sides equal,
  \item the number of hanging symbols cannot be greater than the sum of possible non-terminal symbols on the right side of the remaining rules. If adding a new rule would violate this, it should utilize the necessary number of hanging symbols on its right side. This principle guarantees, that all hanging symbols will be connected. 
\end{itemize}

After all rules have been created, the symbol that was the most recently created is converted into the start symbol. This operation makes all symbols and rules, which are at that moment only productive also achievable and consequently the whole grammar consistent.

The example of the algorithm process was presented in section \ref{teststudy}.

\subsection{Theoretical analysis} \label{tanalysis}
 In this section, a mathematical approach to describe presented previously algorithm was made. Theoretical analysis was performed to provide a better understanding of the system and reveal strict connections between parameters. Those were utilized to shape boundaries and limitations for the system.
 
\subsubsection{Definitions}
Let $S_T$, $S_{NT}$ be the maximum number of terminal and non-terminal symbols, where $S_T$, $S_{NT} \in \mathbb{N}_+$. Numbers of parenthesis rules with non-terminal symbol, parenthesis rules without a non-terminal symbol, iterative rules and branch rules are denoted as $R^+_P$, $R^-_P$, $R_I$ and $R_B$ respectively, where $R^+_P, R_I, R_B\in \mathbb{N}_0$ and $R_B\in \mathbb{N}_+$.

\subsubsection{The ability to generate grammar}
The first insight is the fact that the non-terminal rules do not affect each other in terms of generating rules - i.e., the creation of any $A$ type rule on an existing set of symbols does not exclude the possibility of creating another $B$ type rule. So it is possible to create a grammar composed only of terminal rules and of non-terminal rules of one type. The maximum number of non-terminal rules depends on the number of non-terminal and terminal symbols.

On the other hand, terminal rules force the minimum number of other types of rules - the creation of terminal rules is accompanied by the creation of new non-terminals, which have to be combined using the other rules.

\subsubsection{Parenthesis rules without non-terminal symbols}

This type of rule is defined in a $A \rightarrow ab$ form. It can be created in several ways. It is possible to create a new non-terminal symbol for a rule every time. In such a situation, the number of newly created non-terminals will be equal to the number of available rules, assuming that the maximum number of non-terminal symbols is not exceeded. The second way is to use the same non-terminal symbol on the left - the limit here is the maximum number of combinations that can be created by terminals. When there are no more combinations, it is necessary to create  a new non-terminal.

Adding parenthesis rules without non-terminal symbols produces "hanging" non-terminals - it results from the fact that it is not possible at this time to make both symbols achievable by ensuring the availability of one of the vertices - they must be combined into one tree at a later stage.

Parenthesis rules without non-terminal symbols and with different non-terminals on the left side have to be at most one more than the other rules - otherwise it will not be possible to combine all non-terminals into one tree (the other rules would not be able to connect them).

The number of non-terminal symbols that have to be created when adding parenthesis rules without non-terminal symbols is greater than or equal to (\ref{EQ:1}).

\begin{equation}
\label{EQ:1}  \frac{|R_{P}^{-}|}{S_{T}^{2}}
\end{equation}

This results in given conditions ((\ref{EQ:2}) and (\ref{EQ:3})) for the possibility of generating grammar.

\begin{equation}
\label{EQ:2} \frac{|R_{P}^{-}|}{S_{T}^{2}}\leqslant S_{NT}
\end{equation}
\begin{equation}
\label{EQ:3} \frac{|R_{P}^{-}|}{S_{T}^{2}}\leqslant |R_{P}^{+}|+|R_{I}|+2|R_{B}|+1
\end{equation}

\subsubsection{Parenthesis rules with a non-terminal symbol}

Rules in a form of $A \rightarrow aBb$ are considered as parenthesis rules with a non-terminal symbol. Similarly to the previous case, we can add them in two ways - by using existing non-terminals or by adding new ones.

The rules can be added for individual non-terminal pairs (a non-terminal symbol on the left and a non-terminal symbol on the right), e.g. \emph{(A, B)}. Pairs do not have to be unique - they are two-element variations of the set of non-terminal symbols existing at a given time. The number of variations is therefore $S_{NT}^2$. There can be many rules for a single pair of non-terminals, depending on the number of terminals - in the case of a parenthesis rule with a non-terminal, the maximum number of terminal rules that can be created using a single pair is $S_T^2$.

The number of non-terminals we need to use is greater than or equal to (\ref{EQ:4}).

\begin{equation}
\label{EQ:4} \sqrt{\frac{|R_{P}^{+}|}{S_{T}^{2}}}
\end{equation}

The number of parenthesis rules is limited by the number of rules that we can add or the maximum number of non-terminal symbols. This results in the (\ref{EQ:40}) condition for generating grammar.

\begin{equation}
\label{EQ:40} \sqrt{\frac{|R_{P}^{+}|}{S_{T}^{2}}}\leqslant S_{NT}
\end{equation}

\subsubsection{Iteration rules}
Iteration rules are represented as $A \rightarrow aB$ and $A \rightarrow Ba$. They are very similar to parenthesis rules with a non-terminal symbol from the perspective of generating grammars - they have one non-terminal symbol on the left and one on the right side. They differ by the number of rules that can be created with a single pair of non-terminals - it is equal to $2S_T$. The number of possible non-terminal pairs, however, is the same and equal to $S_{NT}^2$.

The number of non-terminals we need to use to create all iterative rules is therefore greater than or equal to (\ref{EQ:5}).

\begin{equation}
\label{EQ:5} \sqrt{\frac{|R_{I}|}{2S_{T}}}
\end{equation}

The number of iteration rules is limited by the number of rules that we can add or the maximum number of non-terminal symbols. This results in the condition (\ref{EQ:6}) for the possibility of generating grammar.

\begin{equation}
\label{EQ:6} \sqrt{\frac{|R_{I}|}{2S_{T}}}\leqslant S_{NT}
\end{equation}

\subsubsection{Branch rules}

Branch rules are expressed in a form of $A \rightarrow BC$. To create them, we use three-element variations of the set of non-terminal symbols existing at a given time. The number of possible variations is equal to $S_{NT}^3$. These rules do not have terminal symbols, so one "triple" can only be used once.

The number of non-terminals we have to use to create all branch rules is therefore greater than or equal to (\ref{EQ:7}).

\begin{equation}
\label{EQ:7} \sqrt[3]{|R_{B}|}
\end{equation}

The number of branch rules is limited by the number of rules that we can add or by the maximum number of non-terminal symbols. This results in the (\ref{EQ:8}) condition for a grammar to be generated.

\begin{equation}
\label{EQ:8} \sqrt[3]{|R_{B}|}\leqslant S_{NT}
\end{equation}

\subsubsection{Summary}

To sum up, grammar can be generated if the (\ref{EQ:9}) system of inequalities is not contradictory.

\begin{equation}
\label{EQ:9}
\left\{\begin{matrix}
\frac{|R_{P}^{-}|}{S_{T}^{2}}\leqslant S_{NT}\\ 
\frac{|R_{P}^{-}|}{S_{T}^{2}}\leqslant |R_{P}^{+}|+|R_{I}|+2|R_{B}|+1\\
\sqrt{\frac{|R_{P}^{+}|}{S_{T}^{2}}}\leqslant S_{NT}\\
\sqrt{\frac{|R_{I}|}{2S_{T}}}\leqslant S_{NT}\\
\sqrt[3]{|R_{B}|}\leqslant S_{NT}	
\end{matrix}\right.
where
\left\{\begin{matrix}
S_{NT},S_{T},|R_{P}^{-}|\in\mathbb{N}_{+}\\
|R_{P}^{+}|,|R_{I}|,|R_{B}|\in\mathbb{N}_{0}
\end{matrix}\right.
\end{equation}

\subsection{Theoretical justification}
The previously discussed approach allowed to present system properties and limitations in a clear and understandable way. In this section it will be extended with mathematical fundamentals.

\subsubsection{Rule structure analysis}

In the beginning, a deeper insight into the rule structure is needed. Rules can be constructed using two types of symbols - terminal and non-terminal ones. Those types can be presented as a set (\ref{EQ:12}).

\begin{equation}
\label{EQ:12} S_{type} = \left \{ S_N, S_\Sigma \right \}
\end{equation}

where $S_N$ is a non-terminal type and $S_{\Sigma}$  a terminal symbol type. Rules can be differentiated into types too, each type can be linked with a set of symbols of a specific type in a determined order. Context-free rule type can be defined as (\ref{EQ:13}).

\begin{equation}
\label{EQ:13} R^{type} = S_N \times rhs
\end{equation}

where $rhs$ is a \emph{right-hand side} of the rule type, and is a set of symbol types for each position (\ref{EQ:14}).

\begin{equation}
\label{EQ:14} rhs = S_{type}^*
\end{equation}

Rule types can be aggregated in a rule class, each rule class can be described by multiple rule types (\ref{EQ:15}) - e.g. a class of rules with the only non-terminal symbol on the first position on the right-hand side, and any number of terminal symbols.

\begin{equation}
\label{EQ:15} C = \left \{ R_{C_1}^{type}, R_{C_2}^{type},..., R_{C_N}^{type}\right \}
\end{equation}

We denote that any set of rules $R_x$ belong to class $C_x$ by (\ref{EQ:16}).

\begin{equation}
\label{EQ:16} C_x \rightarrow R_x
\end{equation}

Finally, we can also define a set of rule classes for the given grammar $G$ as $C(G)$ (\ref{EQ:17}).

\begin{equation}
\label{EQ:17} C(G) = \left \{ C_1, C_2,...,C_n\right \}
\end{equation}

One additional relation between types has to be defined. We consider rule types as equal if the size of both rule types is the same (number of symbols used in rules) and all symbols on corresponding positions are equal (\ref{EQ:18}).

\begin{equation}
\label{EQ:18}R^{type}_a=R^{type}_b \Leftrightarrow  (|R^{type}_a|=|R^{type}_b|)\wedge \forall_{i\in\left \langle 1, |R^{type}_a| \right \rangle}R^{type}_{ai}=R^{type}_{bi}
\end{equation}

Definitions given above provide methods for classification of structures defined within any context-free grammar and provide additional insights.

Firstly, it is easy to notice that if two classes do not share any rule types, then rule sets that are created within those classes are divergent (\ref{EQ:19}).

\begin{equation}
\label{EQ:19} C_a \cap C_b =\varnothing\Rightarrow R_a \cap R_b=\varnothing
\end{equation}

This insight can be easily extended to entire grammar - if all classes within grammar do not share any rule types, then the number of grammar rules is equal to the sum of rules that belong to each class (\ref{EQ:20}).

\begin{equation}
\label{EQ:20} \forall_{C_a, C_b\in C(G)\land C_a \neq C_b} C_a \cap C_b =\varnothing \Rightarrow |R|= \sum_{i=1}^{|C(G)|} |R_{C(G)_i}|
\end{equation}

\subsubsection{Application for the designed system}

In this section we will take advantage of the previously introduced notation to describe the designed system. Firstly, rule classes, that describes each rule set, have to be defined (\ref{EQ:21}).

\begin{equation}
\label{EQ:21}\begin{array}{l}
C_P^{+} = \left \{ S_N \times S_{\Sigma}S_NS_{\Sigma} \right \}\\
C_P^{-} = \left \{ S_N \times S_{\Sigma}S_{\Sigma}\right \}\\
C_B = \left \{ S_N \times S_{N}S_{N} \right \}\\
C_I = \left \{ S_N \times S_{\Sigma}S_{N}, S_N \times S_{N}S_{\Sigma} \right \}\\
C_P^+ \rightarrow R_P^+ \\
C_P^- \rightarrow R_P^- \\
C_B \rightarrow R_B \\
C_I \rightarrow R_I \\
\end{array}
\end{equation}

The further analysis will be performed for the whole grammar family $(G_{gen})$, that is possible to generate using our method. (\ref{EQ:22}).

\begin{equation}
\label{EQ:22}G_{gen} = \left \{ G_1, G_2,..., G_n\right \}
\end{equation}

All those grammars are built using rules that belong to previously defined classes (\ref{EQ:23}).

\begin{equation}
\label{EQ:23}\forall_{G_i\in G_{gen}} C(G_i)=\left \{ C_P^+,C_P^-,C_B,C_I \right \}
\end{equation}

It is also worth reminding that method parameters for terminal and non-terminal symbols are maximum values (\ref{EQ:24}).

\begin{equation}
\label{EQ:24}\begin{array}{l}
|N|\leq S_{NT}\\
|\Sigma|\leq S_{T}
\end{array}
\end{equation}

Based on the prior knowledge, each rule type can be studied separately. It is easy to notice that the number of rules of each class created during the generation process is lower or equal to the maximum number of possible rules, that could be created using existing terminal and non-terminal symbol sets. This insight, for parenthesis rules with a non-terminal symbol can be expressed as (\ref{EQ:25}).

\begin{equation}
\label{EQ:25}|R_p^+|\leqslant|\left \{ S_N \times S_{\Sigma}S_NS_{\Sigma} \right \}|=|N||\Sigma||N||\Sigma|=|N|^2|\Sigma|^2 \leqslant S_{NT}^2S_T^2
\end{equation}

After collapsing and transforming previous inequality we receive (\ref{EQ:26}).

\begin{equation}
\label{EQ:26}\begin{array}{l}
|R_p^+| \leqslant S_{NT}^2S_T^2\\
\sqrt{\frac{|R_p^+|}{S_T^2}}\leqslant S_{NT}
\end{array}
\end{equation}

The analogous procedure can be applied to another rule classes, like:

\begin{itemize}
\item parenthesis rules without terminal symbol (\ref{EQ:27}).
\begin{equation}
\label{EQ:27}\begin{array}{l}
|R_p^-|\leqslant|\left \{ S_N \times S_{\Sigma}S_{\Sigma} \right \}|=|N||\Sigma||\Sigma|=|N||\Sigma|^2 \leqslant S_{NT}S_T^2\\
|R_p^-| \leqslant S_{NT}S_T^2\\
\frac{|R_p^-|}{S_T^2} \leqslant S_{NT}
\end{array}
\end{equation}

\item iterative rules (\ref{EQ:28}).
\begin{equation}
\label{EQ:28}\begin{array}{l}
|R_I|\leqslant|\left \{ S_N \times S_{\Sigma}S_{N} \right \}\cup \left \{ S_N \times S_{N}S_{\Sigma} \right \}|=\\
|N||\Sigma||N|+|N||N||\Sigma|=2|N|^2|\Sigma| \leqslant 2S_{NT}^2S_T\\
|R_I|\leqslant 2S_{NT}^2S_T\\
\sqrt{\frac{|R_I|}{2S_T}} \leqslant S_{NT}
\end{array}
\end{equation}

\item branch rules (\ref{EQ:29}).
\begin{equation}
\label{EQ:29}\begin{array}{l}
|R_B|\leqslant|\left \{ S_N \times S_{N}S_{N} \right \}|=|N||N||N|=|N|^3\leqslant S_{NT}^3\\
|R_B|\leqslant S_{NT}^3\\
\sqrt[3]{|R_B|}\leqslant S_{NT}
\end{array}
\end{equation}
\end{itemize} 

Another situation that has to be studied is the necessity of connecting all non-terminal symbols created with terminal rules. The only terminal rule type in the developed system are parenthesis rules without the non-terminal symbol. For the purposes of consideration, we assume that we utilize all possible terminal symbols and as few as possible non-terminal symbols to create parenthesis rules without non-terminal symbols on their right-hand side (\ref{EQ:30}).

\begin{equation}
\label{EQ:30}\begin{array}{l}
|R_p^-|\leqslant N_{min}S_T^2\\
\frac{|R_p^-|}{S_T^2}\leqslant N_{min}
\end{array}
\end{equation}

We need to determine how many non-terminal rules is needed to connect non-terminal symbols created with parenthesis rules without-non terminal symbols on their right-hand side. The connection procedure results in the creation of a consistent grammar. The simplest approach is to use one symbol as a root node and connect others using non-terminal rules. It is easy to see, that one rule is able to connect to the root node as many symbols as many non-terminal symbols appear on its right-hand side. The number of non-terminal symbols on a rule of a given class right-hand side can be expressed as (\ref{EQ:31}).

\begin{equation}
\label{EQ:31}rhs(C_i)_{NT}
\end{equation}

Therefore, the total number of symbols possible to connect is the number of each class rules multiplied by the number of non-terminal symbols on the right-hand side of the given rule class increased by the root symbol (\ref{EQ:32}). 

\begin{equation}
\label{EQ:32}N_{min} = 1+\sum_{i=1}^{i=|C(G)|}|R_i|rhs(C_i)_{NT}
\end{equation}

When this equation is applied to the created generation method we obtain (\ref{EQ:33}).

\begin{equation}
\label{EQ:33}\begin{array}{l}
N_{min} = |R_{P}^{+}|+|R_{I}|+2|R_{B}|+1\\
\frac{|R_p^-|}{S_T^2}\leqslant |R_{P}^{+}|+|R_{I}|+2|R_{B}|+1
\end{array}
\end{equation}

All those final equations (\ref{EQ:26})(\ref{EQ:27})(\ref{EQ:28})(\ref{EQ:29})(\ref{EQ:33}) were previously introduced in the equation set (\ref{EQ:9}).

\subsection{Conclusions}
The theoretical analysis revealed dependencies between system parameters, that are vital for algorithm proper output. 

Firstly, the equation set that conditions the possibility of a generation with given parameters was formulated, which allows saying in advance that, while accepted, input parameters put into the algorithm would result in a proper grammar.

Secondly, the general framework that allows an analysis of eventual new rule types and their impact on the algorithm was created. Considering them within the framework would result in new constraints and conditions for proper grammar generation. 

Finally, proper fundamentals for further analysis was made.

\section{Grammar Complexity} \label{gcisec}
There are many approaches to the problem of defining and determining grammar complexity in the literature. 

The most basic was described in \cite{GRUSKA1969152}. At first, the grammatical level of grammar is defined as a maximum set of grammar rules which left-side symbols are mutually dependent. Then generic grammar classification is presented as a non-negative integer. In the following sections of the paper diverse types of classification are presented, such as the number of variables, the number of productions, the number of grammatical levels, the number of non-elementary grammatical levels and the maximum depth of grammatical levels. Every classification and relation between them is meticulously justified and proved mathematically.

In the \cite{gruska1972size} author investigated a more sophisticated criterion of a grammar complexity - the number of all occurrences of all symbols in the rules of grammar. The additional analysis is also performed.

The complexity can be analyzed also from a language perspective. It refers to its origin - what is the minimum number of rules of given grammar type that would be needed to generate a given language? If we denote that type of grammar as $X$, and the language as $L$, then this complexity will be named as $X$ complexity of $L$. This view was considered in \cite{BUCHER1981227}.

Another approach was developed for the use of experimental psychology. Artificial grammars within this area are used to test the ability of subjects to implicitly learn the structure of complex rules. In \cite{Bollt2000} the topological entropy is introduced as the complexity measure. It represents grammar as the sequence space of a dynamical system and calculates complexity by finding the largest eigenvalue of an associated transition matrix. Authors developed appropriate mathematical tools to analyze it and prove its usability as complexity measure of grammar, in particular in the context of implicit learning. Examples are provided as well.

One of the most advanced methods is Omphalos \cite{starkie2004omphalos}. It was introduced for the need of the context-free learning competition Omphalos, which was part of the 7th International Colloquium on Grammatical Inference. Its definition refers to fundamental theorems of the Grammar Inference. The measure was derived by creating a model of the learning task based upon a brute force search (The BruteForceLearner algorithm) to find the characteristic set, which was utilized to determine the demanded measure.

Our solution implements grammar complexity measures in two areas:
\begin{itemize}
  \item obtaining a complexity of already generated grammar
  \item defining a target complexity for generated grammar
\end{itemize}

Obtaining a complexity of already generated grammar with parameters declared explicitly can be utilized to compare many grammars, generated both with our generator and external ones, as they can be imported into our system. We chose Omphalos as a grammar complexity measure for already generated grammar, as it occurs to be the most bound to the grammar inference core theory.

The introduction of target complexity was justified by the need to group generated grammars. In selected real-life applications we do not always need precise control of each rule type number, while it is more important to receive a set of grammars that represents a given complexity, not a precise form. Those sets allow us to perform tests on the intersection of possible grammars within a given complexity, proving algorithms’ universal applicability.

Using Omphalos as target complexity measure was not possible because there is no straightforward algorithm to obtain our parameters or grammar at all using it. The same issue was linked with other algorithms - the reverse process was not described and obtained, mainly because of its potential complexity and ambiguity, but also because of the lack of their authors' research need. However, we made an approach to the theoretical derivation of a method to obtain generation parameters using the number of grammar rules. In the following section, this approach will be described. Output parameters of the given algorithm will be utilized as an input for a grammar generator algorithm.

\subsection{Number of grammar rules - theoretical approach to parameters acquisition}

The algorithm for determining the rules for a given number of grammar rules was designed in the context of the grammar generator presented in this paper. Due to that, parameters that would be selected based on the given rule number are those which the grammar generator would use to create a new grammar. As we previously described, grammar generator parameters can be split into two categories: those which determine the precise properties of the grammar (the number of each rule type) and those which set edge-values (the number of terminal and non-terminal symbols).

The algorithm starts by choosing rule numbers, due to the fact that they, as previously mentioned, determine an exact value for a grammar property. Symbol counts are more flexible, so choosing their values in the next stage would be more feasible. Moreover, assuming that symbol counts would be adjusted to rule numbers, there are only two limitations - the count of parenthesis rules has to be equal to or greater than 1, and all rules have to sum up to the given value. Under those constraints, each rule number is selected randomly. 

In the second stage, maximum numbers of terminal and non-terminal symbols are determined based on previously generated rule numbers. Those numbers have a form of bounds, from which crisp values are randomly selected. Lower bounds have physical meaning - below those values, there would not be a possibility to generate grammar. Upper bounds are recognized as those, above which increasing value does not affect generation process, therefore should not be considered in selection.

Firstly, insights based on the relationship between rules and symbols numbers given in the equation set (\ref{EQ:9}) are needed. If we know the exact number of rules, (\ref{EQ:33}) can be transformed to obtain the lower limit of terminal symbols (\ref{EQ:34}).

\begin{equation}
\label{EQ:34}\sqrt{\frac{|R_{P}^{-}|}{|R_{P}^{+}|+|R_{I}|+2|R_{B}|+1}}\leqslant S_{T}
\end{equation}

As previously discussed, there is no strict upper limit for this parameter since the grammar generator algorithm is not forced to create an exact number of terminal symbols. However, there is upper bound that could be distinguished, which concerns the maximum number of terminal symbols that could be realistically applied to rules. This happens when in each possible position a unique terminal symbol is applied. Therefore the upper limit depends on the number of possible positions created by a potential rule. Those positions are created by rules with terminal symbols on their right-hand side, in this case both types of parenthesis rules. Each type utilizes two terminal symbols, so as an upper limit for the terminal symbol number we can consider (\ref{EQ:35}).

\begin{equation}
\label{EQ:35}S_{T} \leqslant 2|R_{P}^{-}|+2|R_{P}^{+}|
\end{equation}

Based on this data, we can select terminal symbols number from a range (\ref{EQ:36}).

\begin{equation}
\label{EQ:36}S_{T} \in \left \langle \sqrt{\frac{|R_{P}^{-}|}{|R_{P}^{+}|+|R_{I}|+2|R_{B}|+1}}, 2|R_{P}^{-}|+2|R_{P}^{+}|\right \rangle
\end{equation}

The last remaining parameter to be obtained is the non-terminal symbol maximum number. As a result of (\ref{EQ:36}) we can derive the lower limit for this value (\ref{EQ:37}).

\begin{equation}
\label{EQ:37}\frac{\frac{|R_{P}^{-}|}{S_{T}^{2}}+\sqrt{\frac{|R_{P}^{+}|}{S_{T}^{2}}}+\sqrt{\frac{|R_{I}|}{2S_{T}}}+\sqrt[3]{|R_{B}|}}{4}\leqslant S_{NT}
\end{equation}

Again, there is no strict upper limit for this parameter, though there are physical limitations for non-terminal symbols. This limitation appears in the term of a maximum number of non-terminal symbols that could be created during the generation process. During the previously defined generation process, each rule can create at most one non-terminal symbol (since creating a non-terminal symbol is possible during left-hand side symbol selection). However, some left-hand side non-terminal symbols created for parenthesis rules without a non-terminal symbol have to be matched before the procedure ends. Connecting using rule other than branch rule prevent creating a new non-terminal symbol for them. However, a branch rule can still connect one of them while creating a new symbol. As a result, the maximum number of non-terminal symbols is reduced by the difference between the number of branch rules and parenthesis rules without a non-terminal symbol, when there are fewer branch rules. The upper limit for non-terminal rules can be expressed as (\ref{EQ:38}).

\begin{equation}
\label{EQ:38}\left\{\begin{matrix}
S_{NT} \leqslant |R_{P}^{-}|+|R_{P}^{+}|+|R_{I}|+|R_{B}|\ \ for\  \ |R_{B}|\geq |R_{P}^{-}|-1 \\ 
S_{NT} \leqslant |R_{P}^{+}|+|R_{I}|+|2R_{B}|+1 \ \ for\  \ |R_{B}|< |R_{P}^{-}|-1 
\end{matrix}\right.
\end{equation}

Using those limits, we select non-terminal symbols’ maximum value.

At the end of this step we obtain a complete set of random input parameters, yet having a specified number of rules and being able to create a consistent grammar. 

\section{Positive set generator} \label{regpositive}

Creating an artificial grammar was the first step in the process of preparing data for learning benchmarks. The second step is creating a proper test set, both positive and negative ones. The creation of a positive set brings about numerous issues. 

The easiest approach is to enumerate all positive examples: however, this set can be infinite. To avoid that, the example selection is needed. 

The earliest approaches to generate test sentences were presented in \cite{sauder1962general} and \cite{hanford1970automatic}. The first paper was focused on generating test data for Cobol compilers. The second one introduces a method, which allows creating sentences that are syntactically and semantically correct, which permits to produce tests for a complete compiler.

In \cite{Purdom1972} the author focused exclusively on context-free grammars. The algorithm generates a set of short in length examples. They are generated in a way, that ensures, that each production from the grammar is used at least once. Another presented advantage is a short time of execution.

In \cite{ponty2006genrgens} authors present the universal tool GenRGenS, that generates test sets for the various input types. The provided algorithm allows generating combinatorial structures uniformly at random. Since it is widely utilized across many publications within the grammar inference field, It was integrated into our method and described in section \ref{genrgens}.

A suggestion of some extensions for the algorithm applied for GenRGenS was presented in \cite{alain}. Authors considered generating words based on a given distribution of symbols, according to two variants - generation according to exact and expected frequencies. The performed research was motivated by a possible application in genomics.

Another approach claims, that the selection has to be handled in an adequate manner - a set should represent every aspect of the grammar. One of the solutions to this task was provided in the paper \cite{optimalTestSets} which describes the algorithm, that creates an optimum test set of maximum size $2|R|^3$. The detailed description will be presented in section \ref{optimal}. Due to the proved effective representation of the source grammar, this solution was integrated into our method.

\subsection{Optimal Test Sets for Context-Free Languages} \label{optimal}

\begin{figure}[ht!]
\begin{center}
\includegraphics{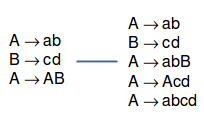}
\caption{An example of conversion to a linear grammar}
\label{lineargrammar}
\end{center}
\end{figure}

The algorithm starts with the conversion of an input grammar to a linear grammar (Fig. ~\ref{lineargrammar}).

\begin{figure}[ht!]
\begin{center}
\includegraphics{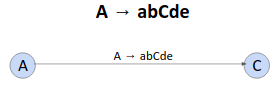}
\caption{An linear grammar graph edge}
\label{linearedges}
\end{center}
\end{figure}

Then a graph based on the previously converted linear grammar is created. In this graph, nodes represent non-terminal symbols and the string-ending symbol $\Gamma$. Edges are created based on rules and are attached from non-terminal symbols on their left-hand sides to non-terminal symbols on their right-hand sides. The edges are labeled with terminal symbols on their right-hand side at the associated rule (Fig.~\ref{linearedges}).

\begin{figure}[ht!]
\begin{center}
\includegraphics{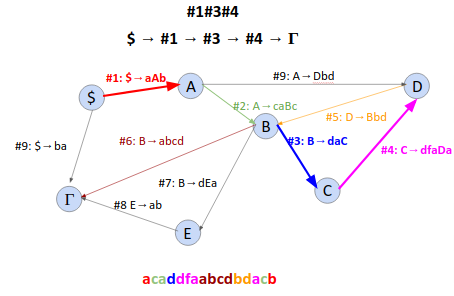}
\caption{A linear grammar graph}
\label{lineargraph}
\end{center}
\end{figure}

The graph which is built is the basis for generating positive examples. To do this, a set that contains every one-, two- and three-element combinations of linear grammar rules is created. For each combination path that starts in the node, represents the start symbol, ends in $\Gamma$ and contains all edges related to the rules in combination in a given order is determined. If an example is created by following this path, it is added to the positive test set (Fig. ~\ref{lineargraph}).

\subsection{GenRGenS: Generation of Random Genomic Sequences and Structures} \label{genrgens}

GenRGenS is a software tool that allows users to randomly generate genomic sequences and structures. It supports different sequence analysis approaches like Markov Chains, hidden Markov models, weighted context-free grammars, regular expressions and PROSITE expressions and use them to model and generate structured objects.

For our generator, we integrated GenRGenS in a WCFG mode. The algorithm can generate words of a given length that obey the grammar, such that the probability of any generated word is proportional to its weight. Generation is based on algorithm \cite{flajolet1994calculus}, that allows generating combinatorial structures uniformly at random.

\subsection{Methods comparison}
The algorithm described in \cite{optimalTestSets} due to the solid theoretical background appears to be more reliable and productive in terms of application as input data for grammar induction - it promises coverage of all language aspects (proved mathematically). Meanwhile, random selection with GenRGenS, although unified, does not guarantee that all properties of source grammar will be mapped on the example set.

Both approaches were also compared at the level of practical implementation. We observed that the Optimal Sets algorithm tends strongly to minimize the number of examples, especially for small grammar, what appears to become an issue for the grammar inference. However, for more complex grammars it provides a solid example set. In comparison, getting an adequate number of examples using GenRGenS requires multiple runs with different parameters, because during one run it generates examples only of a given length. The output of multiple runs is summed to one set.

Regardless of given insights, both of them were applied for the sake of data diversity.

\section{Negative set generator} \label{regnegative}

Negative examples for the given grammar can be obtained in multiple ways. The most basic approach is a generation of random sequences. More sophisticated ones involve modifying previously created positive examples. All of them requires checking if created examples are not parsed by the origin grammar.  Classification of the generated examples is considered using \emph{CYK} (\emph{Cocke–Yonger–Kasami}) algorithm. It allows examples to be parsed using \emph{CFG} grammar expressed in \emph{Chomsky Normal Form} \cite{younger1967recognition}. Generated grammar is not represented in this form, so it has to be converted. 

Conversion can be achieved using many approaches. The algorithm applied in our system follows the one described in \cite{hopc} with transformation names from \cite{lange2009cnf}. It consists of a sequence of simple \emph{START}, \emph{TERM}, \emph{BIN}, \emph{DEL}, \emph{UNIT}  transformations. \emph{START} is the elimination of the start symbol from the right-hand side step; \emph{TERM} eliminates rules with nonsolitary terminals on the rules’ right sides; \emph{BIN} transforms rules with right-hand sides containing more than 2 nonterminals; \emph{DEL} removes $\varepsilon$-rules; and finally \emph{UNIT} converts unit rules.

Not all of those transformations are necessary for converting grammars described in this paper because a less strict variant of \emph{CNF} (start symbol allowed on the right side of the rule) is applied and a prior form of input rules for the conversion is specified. This allows simple conversion patterns to be created for each rule type.

Branch rules (e.g, $A \rightarrow CD$) do not need to be transformed, because they are already in \emph{CNF} - they always have two nonterminals on their right-hand side.

\begin{figure}[ht!]
\begin{center}
\includegraphics{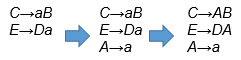}
\caption{An iterative rules conversion example}
\label{iterativeconv}
\end{center}
\end{figure}

Both variants of iteration rules ($A \rightarrow cE$,$A \rightarrow Ec$) are converted in the same way - by replacing a terminal symbol with a nonterminal symbol that corresponds to them and a terminal \emph{CNF} rule. Only one nonterminal symbol and terminal rule replacing a terminal symbol can exist. An example of a conversion of two iterative rules is given in (Fig. ~\ref{iterativeconv}).

\begin{figure}[ht!]
\begin{center}
\includegraphics{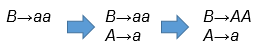}
\caption{An parenthesis rules without non-terminal symbol conversion example}
\label{parenthesisconv}
\end{center}
\end{figure}

To convert parenthesis rules ($A \rightarrow aBb$,$A \rightarrow ab$) properly, we need to prepare a different approach for each variant. A simpler one, $A \rightarrow ab$, demands steps analogous to those with iteration rules - replacing terminal symbol with a nonterminal symbol that corresponds to it. As before, we create new symbols and rules only if needed (Fig. ~\ref{parenthesisconv}).

\begin{figure}[ht!]
\begin{center}
\includegraphics{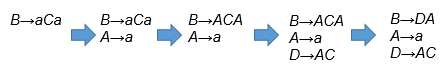}
\caption{An parenthesis rules with non-terminal symbol conversion example}
\label{parenthesiswithconv}
\end{center}
\end{figure}

The conversion of the second variant ($A \rightarrow aBb$) is a little bit more complex. We by replacing terminal symbols, but we end with three nonterminal symbols on the right-hand side, which need to be reduced. We can do this by clustering two of them into one new symbol which is produced using new rule. This new symbol and rule cannot be applied anywhere else to preserve grammar consistency \ref{parenthesiswithconv}).

The number of rules and symbols after conversion can easily be determined. Each terminal symbol produces one new, dedicated nonterminal symbol and rule. Each parenthesis rule with a nonterminal symbol creates an additional nonterminal symbol and rule. New values can be expressed with the following equations (\ref{EQ:39}).

\begin{equation}
\label{EQ:39}\begin{array}{l}
S^{*} += S_T\\
|R^*| += S_T+|R^+_P|
\end{array}
\end{equation}

\subsection{Random sequences generator}
The first step of random sequences generation is user parameter analysis. The negative set is not generated if the generation step flag is not set. The remaining parameters are self-explanatory and describe the negative set examples’ attributes - starting from set cardinality to individual examples’ length.

The next stage is example generation. The algorithm generates a random string of a random length in the range set by the user using terminal symbols and tries to classify it. If the created example is not parsed it cannot be derived using the generated grammar, and therefore it is considered a member of the negative set. If the example is positive, the sequence is re-drawn. This procedure is repeated until the demanded number of negative examples is generated.

\subsection{Levenshtein distance-based negative set generator}

Random sequences, which generation was described in the previous section, can be very different from examples, that can be obtained using given grammar, therefore differentiation between them and positive ones may be found less challenging for grammar inference algorithms.

Another approach takes into account a modification of currently existing positives examples, where the level of change is precisely controlled. Difference between an original example and a modified one is measured with the Levenshtein distance \cite{levenshtein1966binary}. The Levenshtein distance between the two examples is the minimum number of single-character edits (insertions, deletions or substitutions) required to change one example into the other.

The generation process proceeds similarly as during random sequences generation. Firstly, the input parameters are processed. Besides generation flag and the example set cardinality, there is Levenshtein distance between source and target example.

Secondly, a proper generation stage proceeds. At its beginning, a positive source example is chosen. Then, a target example, which Levenshtein distance between it and the source example is equal to that given as a parameter is generated. Before adding to the set, it is verified by \emph{CYK} if it is not parsed by the generated grammar. Generation process repeats until the demanded set cardinality is obtained.

\section{Overall system flow} \label{reqflow}

\subsection{System requirements}

The basis of every algorithm creation is specifying the requirements that it will fulfill. We require that those will be provided as the algorithm input. We highlighted given parameters:
\begin{enumerate}
\renewcommand{\labelenumi}{(\alph{enumi})}
  \item Grammar structure requirements, given as number of all or, alternatively, specific number of each rule type - parenthesis (with and without non-terminal symbol), bracket and iteration rules, with maximal number of terminal and non-terminal symbols,
  \item Positive Set requirements:
   \begin{itemize}
  \item Generation flag for the Minimal Positive Set,
  \item Generation flag for the GenRGenS Posititve test set with maximum number of examples of each length and maximum length of examples.
\end{itemize}
  \item Negative set requirements:
\begin{itemize}
  \item Generation flag for random negative test set generation with number of examples to create and their minimum and maxium length,
  \item Generation flag for Levenstein negative test set generation, number of examples to create and distance between created and original example. 
\end{itemize}
\end{enumerate}

An explanation of all parameters will be contained in the next few sections.

At its output we expect to obtain:
\begin{itemize}
  \item Generated grammar fulfilling given requirements,
  \item Generated grammar converted to Chomsky Normal Form (CNF),
  \item Positive example sets,
  \item Negative example sets.
\end{itemize}

\subsection{Flow}

\begin{figure}[ht!]
\begin{center}
\includegraphics[width=0.8\linewidth]{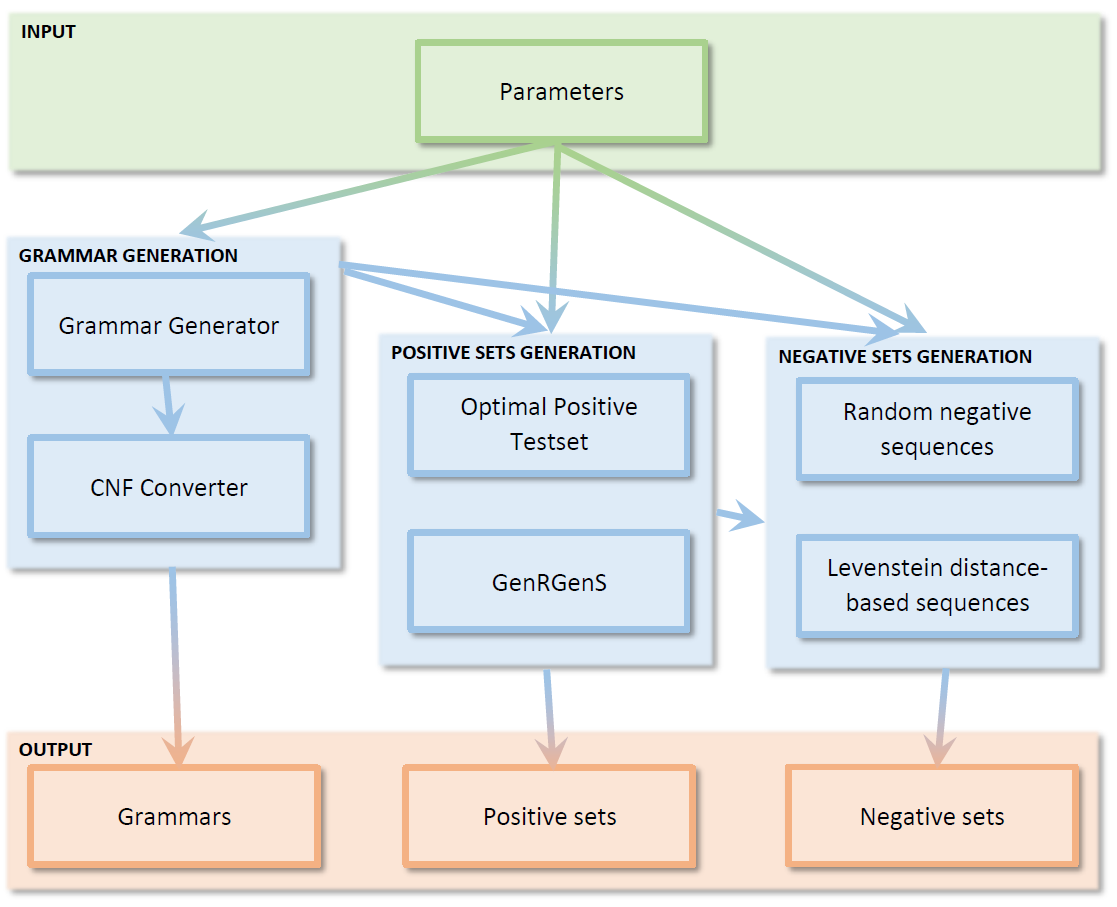}
\caption{Application flow}
\label{flow-figure}
\end{center}
\end{figure}

The application flow could be easily presented with a visual flow chart (Fig.~\ref{flow-figure}).

The algorithm starts with a set of given parameters, which constitute the base to generate an artificial context-free grammar. This grammar, converted to Chomsky Normal Form, is used to create positive and negative example sets. All this data collects into a complete kit for testing grammar induction algorithms performance.

\section{Test Study} \label{teststudy}

Test study was performed for exemplary research need of grammar generation with the rule number equal to 5.

At first, precise parameters should be determined. The algorithm, that allows to obtain them was described in section \ref{gcisec}. The initial step of it is a random selection of rules. An exemplary set of rules could present as below:

\begin{verbatim}
Parenthesis rules without non-terminal symbol - 2
Parenthesis rules with non-terminal symbol - 0
Branch rules - 1
Iterative rules - 2
\end{verbatim}

The number of rules sums up to the demanded value.

Next step is the selection of maximum numbers for non-terminal symbols and terminal symbols. According to equations (\ref{EQ:34}) and (\ref{EQ:35}) number of maximal terminal symbols should be between 1 and 4. Exemplary random value for maximal terminal symbols number is 3. Using this value maximum non-terminal symbols number can be determined. According to equations (\ref{EQ:37}) and (\ref{EQ:38}) it should be placed in range between 1 and 5. Exemplary value from this range is 4. The parameters set was filled with:

\begin{verbatim}
Maximum number of terminal symbols - 3
Maximum number of non-terminal symbols - 4
\end{verbatim}

At this point exact parameters for a given rule number are determined. In the following phase, grammar will be generated. All steps are visualized in the Table~\ref{generationflow}, all recently added or modified are marked in bold.

Grammar generation starts with adding parenthesis rules without a non-terminal symbol. The first rule (Step 1) has to create at least one new non-terminal symbol and one terminal since none of the symbols exist in this step. The second terminal could be either a new one or previously created - a new one is randomly chosen to be created. Rule $A \rightarrow ab$ is added.

In the second step, we still need to create a parenthesis rule without a non-terminal symbol. There is still space for new non-terminal symbols, and for this rule, one is created. For the right-hand side, one terminal symbol is reused, and one is created. Rule $B \rightarrow bc$ is added.

In the third step, there are no parenthesis rules without non-terminal left, so one from remaining types can be drawn - in this step branch rule was selected. It created a new non-terminal symbol and used  the same non-terminals twice on the right-hand side. Rule $C \rightarrow AA$ is added.

The fourth step brings two iterative rules left to create. Random creation uses all previously existing symbols, although there still is space for new non-terminals and terminals. Rule $A \rightarrow Cc$ is added.

In the fifth step, there is only one iterative rule to add. One non-terminal symbol created with parenthesis rule has not been attached yet, so it is forced to be placed on the right-hand side of a new rule. The left-hand side has to be reused from existing ones too. 

The algorithm selects also existing terminal as a terminal for the right-hand side. Rule $C \rightarrow Bc$ is added.

The last step is performed when all demanded rules were created. A recently added non-terminal symbol is transformed into start-symbol, what makes all, currently productive, rules and symbols achievable.

\begin{table}[ht!bp]\small
\caption{The grammar generation flow}
\label{generationflow}
\centering
\begin{tabular}{cc|cc|cc|cc|cc|cc}
\multicolumn{2}{c|}{\textbf{Step 1}} & \multicolumn{2}{c|}{\textbf{Step2}} & \multicolumn{2}{c|}{\textbf{Step 3}} & \multicolumn{2}{c|}{\textbf{Step 4}} & \multicolumn{2}{c|}{\textbf{Step 5}} & \multicolumn{2}{c}{\textbf{Step 6}} \\ \hline
\multicolumn{1}{c|}{\textbf{S}} & \textbf{R} & \multicolumn{1}{c|}{\textbf{S}} & \textbf{R} & \multicolumn{1}{c|}{\textbf{S}} & \textbf{R} & \multicolumn{1}{c|}{\textbf{S}} & \textbf{R} & \multicolumn{1}{c|}{\textbf{S}} & \textbf{R} & \multicolumn{1}{c|}{\textbf{S}} & \textbf{R} \\ \hline
\textbf{A} & $\;$\textbf{A$\rightarrow$ab}$\;$ & A & $\;$A$\rightarrow$ ab$\;$ & A & $\;$A$\rightarrow$ab$\;$ & A & $\;$A$\rightarrow$ab$\;$ & A & $\;$A$\rightarrow$ab$\;$ & A & $\;$A$\rightarrow$ab$\;$ \\
\textbf{a} &  & \textbf{B} & \textbf{B$\rightarrow$bc} & B & B$\rightarrow$bc & B & B$\rightarrow$bc & B & B$\rightarrow$bc & B & B$\rightarrow$bc \\
\textbf{b} &  & a &  & \textbf{C} & \textbf{C$\rightarrow$AA} & C & C$\rightarrow$AA & C & C$\rightarrow$AA & \textbf{\$} & \$$\rightarrow$AA \\
 &  & b &  & a &  & a & \textbf{A$\rightarrow$Cc} & a & A$\rightarrow$Cc & a & A$\rightarrow$\$c \\
 &  & \textbf{c} &  & b &  & b &  & b & \textbf{C$\rightarrow$Bc} & b & \$$\rightarrow$Bc \\
 &  &  &  & c &  & c &  & c &  & c & 
\end{tabular}
\end{table}

\begin{figure}[ht!]
\begin{center}
\includegraphics[width=0.8\linewidth]{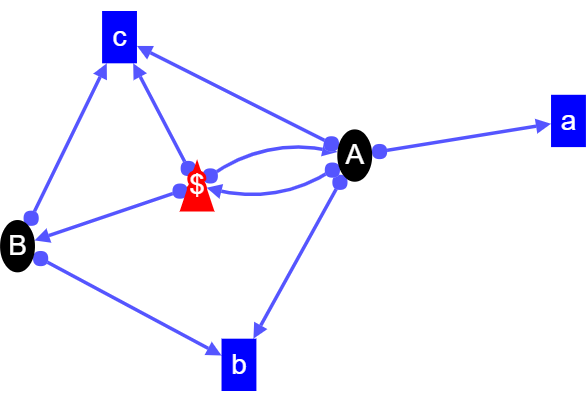}
\caption{The generated grammar visualization}
\label{grammarviz}
\end{center}
\end{figure}

Visualization of the grammar obtained is presented (Fig. ~\ref{grammarviz}). As we see, as a result of the process we obtained functional and consistent grammar.

\section{Implementation} \label{tooldesc}
\begin{figure}[ht!]
\begin{center}
\includegraphics[width=0.8\linewidth]{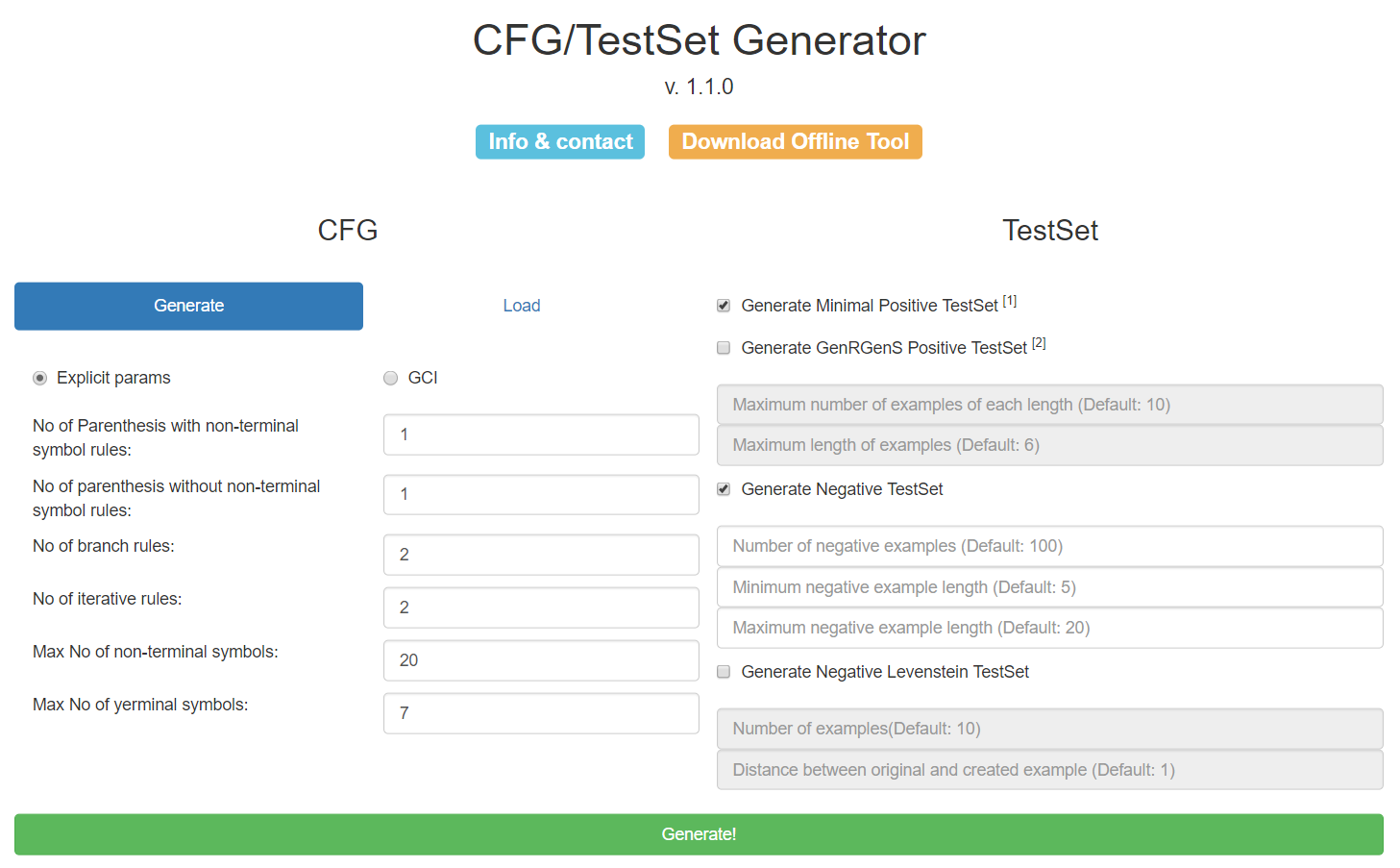}
\caption{The generator main screen}
\label{mainscreen}
\end{center}
\end{figure}

To provide the highest availability for potential users method implementation was designed as online service and is available as a web page via any Internet browser \cite{LanguageGenerator}. Preview of the webpage is presented in (Fig. \ref{mainscreen}). It consists of two main parts.

First one is responsible for generating grammar itself. Grammar can be generated using precise numbers of each rule type or with setting a number of all rules. Alternatively, previously generated grammar can be loaded.

The second one is described as a TestSet part. It allows selecting types of sets that should be generated for the grammar obtained in a way described in the previous part. Parameters for all methods can be also selected.

Buttons that allow to navigate to the informational page or downloading offline implementation are also available. Button, that starts the process of generation is present at the bottom of the page.

As the grammar and test sets are generated, the results screen is presented. The main part of it is tab section, that contains information about originally generated (or loaded) grammar, grammar converted to CNF and created test sets.

Both grammars are described with a list of symbols and rules, that they consist of, and graphical representation. There is also a button, that allows to download them.

Test sets tab presents generated sets, with a list of created examples and possibility to download them as \emph{*.txt} files.

The top of the screen displays field with a sharable link to the current report and the indicator with a determined grammar complexity.

The detailed guide was also published online, as a reference from the main page.

\section{Discussion} \label{conc}

Grammars with both positive and negative test sets, that were created using the presented method were utilized as an input for a grammar inference system - GCS. The output observation proved assumptions claimed in section \ref{introduct}. We observed the existence of learning performance curve during inference with an increasing number of rules within the generated grammar. We also noticed different performance linked with various grammar constituents. Obtained insights would be put under scrutiny during further research and aid GCS improvement. It is worth to notice, that for grammar, that we pay special attention to (due to its properties) we can generate new grammars, that are similar in terms of those properties. It could proceed at any stage of the research.

In general, obtained results suggest, that generated grammars could be also widely used across the grammar inference field and hopefully would give other researchers useful tool for verifying the performance of designed algorithms and makes a comparison of results, for a given grammar class, possible.

The details analysis of numerous generated grammars was also performed. It was noticed that generated grammars have, on average, more symbols that are placed closer to terminal symbols than start symbol. It is a result of our generation method - symbols, that were created earlier during the generation process are more likely to be connected due to the higher number of draws, in which they could appear. The observed property requires further research to make generation process more uniform.

\section{Future Work} \label{fwork}
Future work will focus on two different aspects of our system - one part connected with technical and scientific issues and other parts that affect sharing grammar inference knowledge with the community.

Strictly scientific research will focus around introducing new grammar attributes, that describes their behavior in terms of example generation and structure. Consequently, that would lead to new parameters creation for the generation process, allowing generated grammar specific attributes to be easier to control and customize.

Implementation of new types of grammars to generate (eg. context grammars) is also considered in the following phases, however, it demands deep prior research for a theoretical background.

Community development goals are much more clear - it is planned to create a community-driven database of artificial grammars sets, each tagged by specific attribute (eg. set of grammars aimed to test the performance of detecting specific structures) with benign graduation of complexity. Community members would be allowed to add their own sets, examination results and insights. Those facilities are meant to boost research focused on specific problems and make a comparison of numerous algorithms easily accessible.

\nocite{*}
\bibliographystyle{fundam}
\bibliography{iterative}


\end{document}